\documentclass[prl,showpacs,preprintnumbers,amsmath,amssymb, longbibliography, superscriptaddress,nofootinbib,twocolumn]{revtex4-1}
\usepackage{booktabs}
\usepackage{epsfig}
\usepackage{graphicx}
\usepackage{color}
\usepackage{shuffle}
\usepackage{hyperref}
\usepackage{wrapfig}
\usepackage{cases}
\usepackage[utf8]{inputenc}
\usepackage{mathtools,tabu}
\usepackage{slashed}
\usepackage{empheq}

\usepackage{stackengine}
\stackMath

\newcommand{\pcub}{\sigma_3}
\newcommand{\psq}{\sigma_2}

\renewcommand{\ap}{{\alpha'}}
\newcommand{\stAbiAdj}{ s t A^{\rm bi}(s,t) }
\newcommand{\Aym}{A^{\rm YM}}
\newcommand{\Abi}{{A}^{\rm bi}}
\newcommand{\AHD}{A^{\rm HD}}
\newcommand{\fullAym}{{\cal A}^{\rm YM}}
\newcommand{\fullAGR}{{\cal A}^{\rm GR}}

\newcommand{\trivialAXY}{A^{XY}}
\newcommand{\AcssXY}{ A^{XY}_{\rm ss} }
\newcommand{\AcssOO}{ A^{00}_{\rm ss} }
\newcommand{\permAXY}{A^{XY}_{\rm d}}

\newcommand{\Zbi}{Z^{\rm bi}}
\newcommand{\Zsym}{Z_{\rm sym}}
\newcommand{\Zss}{Z^{\rm ss}}
\newcommand{\Zadj}{Z_{\rm adj}}

\newcommand{\jss}{j^{\rm ss}}
\newcommand{\jnl}{j^{\rm nl}}
\newcommand{\dabcd}{d^{abcd}}

\newcommand{\nym}{n^{\rm YM}}
\newcommand{\tnym}{\tilde{n}^{\text{YM}}}

\newcommand{\ctrivXY}{\hat{c}^{XY}}
\newcommand{\cssXY}{\hat{c}^{\,{\rm ss},XY}}
\newcommand{\cpermXY}{\hat{c}^{\,{\rm d},XY}}

\def\be{\begin{equation}}
\def\ee{\end{equation}}
\newcommand\bea{\begin{align}}
\newcommand\eea{\end{align}}
\def\nn{\nonumber}
\def\eqn#1{Equation~(\ref{#1})}
\def\Eqn#1{Equation~(\ref{#1})}
\def\eqns#1#2{Eqs.~(\ref{#1}) and~(\ref{#2})}
\def\Eqns#1#2{Eqs.~(\ref{#1}) and~(\ref{#2})}

\def\Tab#1{Tab.~{\ref{#1}}}
\def\str{\,{\rm symTr}\,}

\AtBeginDocument{
\heavyrulewidth=.08em
\lightrulewidth=.05em
\cmidrulewidth=.03em
\belowrulesep=.65ex
\belowbottomsep=0pt
\aboverulesep=.4ex
\abovetopsep=0pt
\cmidrulesep=\doublerulesep
\cmidrulekern=.5em
\defaultaddspace=.5em
}

\begin{document}
\title{Simple encoding of higher-derivative gauge and gravity counterterms}
\author{John Joseph M. Carrasco}
\affiliation{Department of Physics \& Astronomy, Northwestern University, Evanston, Illinois 60208, USA}
\affiliation{Institut de Physique Theorique, Universite Paris Saclay, CEA, CNRS, F-91191 Gif-sur-Yvette, France}
\author{Laurentiu Rodina}
\affiliation{Institut de Physique Theorique, Universite Paris Saclay, CEA, CNRS, F-91191 Gif-sur-Yvette, France}
\author{Zanpeng Yin}
\affiliation{Department of Physics \& Astronomy, Northwestern University, Evanston, Illinois 60208, USA}
\author{Suna Zekioğlu}
\affiliation{Department of Physics \& Astronomy, Northwestern University, Evanston, Illinois 60208, USA}

\begin{abstract} 
Invoking increasingly higher dimension operators to encode novel UV physics in effective gauge and gravity theories traditionally means working with increasingly more finicky and difficult expressions.  We find that the duality between color and kinematics provides a powerful tool towards drastic simplification. Local higher-derivative gauge and gravity operators at four-points can be absorbed into simpler higher-derivative corrections to scalar theories, requiring only a small number of building blocks to generate gauge and gravity four-point amplitudes to all orders in mass dimension.
\end{abstract}
\preprint{NUHEP-TH/19-13}

\maketitle
Gravitational quantum scattering amplitudes---the invariant quantum evolution of what distance means in space and time, consistent in the classical limit with Einstein's general relativity from the Einstein-Hilbert action (GR)---are much simpler than expected. This simplicity can be traced to the fact that these perturbative graviton dynamics are completely encoded~\cite{KLT,BCJ,BCJLoop, BCJreview} in the predictions of much simpler gluonic or gauge theories in a framework called double-copy construction.

It is currently an open question as to whether any four-dimensional (pointlike) quantum field theory of gravity is perturbatively finite. The most promising case, maximally supersymmetric supergravity, is a subject of much current research exploiting color-kinematics duality and double-copy construction~\cite{ BCJLoop, SimplifyingBCJ,N46Sugra, N46Sugra2, Bern:2012cd,
  Bern:2012gh, Boels:2012sy, Bern:2013qca, Bern:2014lha, UVFiveLoops,
  Herrmann:2016qea, HerrmannTrnkaUVGrav}. While Yang-Mills (YM) theory is famously renormalizable in four dimensions, it ceases to be in higher-dimensions, requiring new physics at short distances.  Independent of any aesthetic inclination towards perturbative finiteness, from an effective field theory perspective, any new higher energy physics will be encoded in the low-energy theory as Wilson coefficients of higher derivative operators, motivating an understanding of what we can definitively clarify about such predictions to all orders in mass dimension.
      
Recent work has shown that at tree-level both the supersymmetric and bosonic open string amplitudes admit field theory double-copy descriptions~\cite{Broedel2013tta,Huang:2016tag,Carrasco2016ldy,Mafra2016mcc,Carrasco2016ygv,Azevedo2018dgo}, pulling the higher-derivative corrections to a putative effective scalar bi-colored theory, encapsulating all order $\alpha'$ corrections, called Z-theory. Inspired by the existence of Z-theory amplitudes, here we consider a bootstrap approach, asking simply what predictions are consistent with unitarity, double-copy structure, gauge invariance, and locality.  We will see the power of striating Bose-symmetric amplitudes into a small number of simple color-dual building blocks allows us to reach all orders of higher-derivatives through elementary considerations. 

Focusing on four-point tree level, we will identify all the single-trace consistent modifications to gauge theory compatible with supersymmetry in terms of three higher-derivative scalar building blocks that, when double-copied with Yang-Mills, reproduce the open superstring. We further identify the four additional independent higher-derivative vector building blocks ($[F^3]$, $[(F^3)^2 + F^4]$,  $[D^2F^4]$, and $[(D F)^4]$), out of the seven potential tensor structures~\cite{Bern:2017tuc}, required to capture the bosonic open string. Related gravity corrections are simply generated by double-copy construction, replacing color factors with color-dual building blocks.

\section{Adjoint-type building blocks.} We will briefly review color-kinematic representations in the adjoint at four-points. We refer the interested reader to Ref.~\cite{BCJreview} for a detailed treatment.  Yang-Mills amplitudes can be expressed in terms of cubic (trivalent) graphs
\be
{\cal A}^{\rm YM}_4= \frac{c_s \nym_s}{s} + \frac{c_t \nym_t}{t} + \frac{c_u \nym_u}{u}\, 
\label{fullYMAmp}
\ee
where $s,t,u$ are four-point momentum invariants following an all outgoing convention as $s=(k_1+k_2)^2$, $t=(k_2+k_3)^2$, and
$u=-s-t$.  We can identify each graph with a distinct ordered list of these invariants, and can thus label both graph weights $c_g$ and $\nym_g$ in terms of each graph's unique list, e.g. $j_s=j(s,t,u)$, $j_t=j(t,s,u)$, and $j_u=j(u,t,s)$.  
The color-weights $c_g$ are simple dressings of adjoint color-generators $f^{abc}$ with repeated indices summed over, and the kinematic weights $\nym_g$ are Lorentz products between external momenta and polarization vectors.  We emphasize that both the color and kinematic weights satisfy Jacobi identities and antisymmetry around vertex flips:
\begin{align}
c_s &= c_t + c_u & c(a,b,c)&=-c(a,c,b) \nn \\
\nym_s &= \nym_t + \nym_u & \nym(a,b,c)&=-\nym(a,c,b)\,.
\label{jacobiRelations}
\end{align}
As such, this is called a color-dual representation of Yang-Mills.  We refer to functions that satisfy both antisymmetry and Jacobi-identities as {\em adjoint-type}, and this representation of Yang-Mills as specifically manifesting an adjoint-type double-copy structure.  

Gauge invariance is maintained by the fact that the color-weights, $c_g$, satisfy anti-symmetry and Jacobi identities. As per double-copy construction, we can replace the adjoint-type color weights with adjoint-type kinematic weights $\tilde{n}_g$ to generate gravity amplitudes invariant under linearized diffeomorphism:
\be
\fullAGR_4=  \frac{\tnym_s \nym_s}{s} + \frac{\tnym_t \nym_t}{t} + \frac{\tnym_u \nym_u}{u}\,.
\label{adjointGR}
\ee
Details of state identification for a variety of (super)-gravity theories can be found in Ref.~\cite{BCJreview}. It is important to realize that the kinematic weights $n_g$ and $\tilde{n}_g$ need not come from the same theory, and that the double-copy construction promotes any global supersymmetry of the kinematic weights into a local supersymmetry of the gravitational amplitude.  As Yang-Mills amplitudes and numerators can be promoted to super Yang-Mills amplitudes in an on-shell superspace we do not distinguish between Yang-Mills and super-Yang-Mills numerators here. 

It will simplify our discussion to introduce the gauge-invariant objects called ordered amplitudes.  Let us cast the color-weights in \Eqn{fullYMAmp} to a minimal basis using \Eqn{jacobiRelations},  by eliminating $c_t$ in favor of $c_s$ and $c_u$. Their coefficients, labeled $\Aym$, are called ordered or partial amplitudes:
\begin{align}
\fullAym_4 &= c_s \left ( \frac{\nym_s}{s} + \frac{\nym_t}{t} \right) + c_u \left(\frac{\nym_u}{u} - \frac{\nym_t}{t} \right)\\
&= c_s \Aym(s,t) + c_u \Aym(u,t)\,.
\label{orderedAmps}
\end{align}
As the $\Aym(a,b)$ appear in the full amplitude with independent color basis elements, they must each be gauge invariant. Expressing these ordered amplitudes in a basis of kinematic weights $n_g$, say by eliminating $n_u$ via \Eqn{jacobiRelations}, demonstrates that the distinct color orders are intimately related.  Indeed one can immediately identify the permutation-invariant quantity $s t \Aym(s,t)=(s t u) \Aym(s,t)/u$.  This discussion carries through for any adjoint-type double-copy amplitude.

We will first be concerned with how we can construct higher-derivative corrections to Yang-Mills by only modifying the color-weights in a manner consistent with its adjoint-type structure. We begin by recognizing that the trivial modification of the color-weights with simple products of permutation-invariant scalar combinations remains adjoint-type:
\be
\boxed{ \ctrivXY_g = \pcub^X \psq^Y c_g  \ap^{3X+2Y}}\,,
 \label{trivialColor}
\ee
where we introduce  (spanning) permutation-invariant factors $\pcub\equiv(stu)$, $\psq\equiv (s^2+t^2+u^2)$, and a dimensional parameter $\ap$ to track mass-dimension.
This results in an ordered $s$-$t$ channel scattering contribution proportional to:
\be
\trivialAXY(s,t) =   \pcub^X \psq^Y \left(\frac{c_s}{s}+\frac{c_t}{t}\right)= \pcub^X \psq^Y  \Abi(s,t)
 \label{trivialAmp1} 
\ee
As any $X$ and $Y$ constitute permutation-invariant scalings of the  ordered amplitude $\Abi(s,t)$ for bi-adjoint scalars, all field theory relations are automatically preserved.  It should be apparent from \eqn{trivialAmp1} that for local corrections, one would restrict to  $X\ge1$. 
 
We now introduce the notion of a Jacobi-identity satisfying composition. Given functional adjoint-type maps $j(a,b,c)$ and $k(a,b,c)$,  we can build an adjoint-type map $n(a,b,c)$ as the composition of $j$ and $k$ via
\be
n_s = {\cal J}(j,k) \equiv  j_t k_t - j_u k_u\,.
\ee
We will use this composition to build a ladder of color-dual scalar weights starting from one linear in Mandelstam invariants, the so called {\em simple scalar} numerator:
\be
\jss(a,b,c) = c-b \,.
\label{jssdef}
\ee
This corresponds to a scalar charged in the adjoint mediated by a massless vector, e.g. with interaction term $f^{abc}A^\mu (\partial_\mu \phi) \phi$. Note that we may rewrite \Eqn{trivialAmp1} in terms of these simple scalar numerators:
\be
 s t \trivialAXY(s,t) =   \pcub^X \psq^Y \left({c_s  \jss_s  + c_t  \jss_t + c_u  \jss_u}\right)
 \label{trivialAmp} 
\ee

What happens when we compose the simple scalar with itself?  We find the adjoint-type kinematic weight associated with the non-linear sigma model,
\be
 \jnl_s= s (u-t) = s \jss_s\propto{\cal J}(\jss, \jss) \,.
 \label{jnlsmdef}
\ee
Any further compositions between $\jss$ and $\jnl$ only differ from these building blocks by powers of $\sigma_2$ and $\sigma_3$---our ladder of scalar numerators closes under products of permutation invariants after two rungs.  

We are ready to consider compositions involving both color weights and kinematic weights: namely, our two distinct scalar numerators.   Only $ {\cal J}(c,\jss)$, is not redundant with the amplitudes provided by \eqns{trivialColor}{trivialAmp},
\be
\boxed{ \cssXY_s = \pcub^X \psq^Y  {\cal J}(c,\jss) \ap^{1+3X+2Y}}\, .
\label{colorAndKinJac}
\ee
Such weights result in ordered $(s,t)$ channel scattering amplitudes proportional to:
\begin{multline}
{s t} \AcssXY(s,t)  
= \pcub^X \psq^Y 
\left( {c_s  \jnl_s+  c_t  \jnl_t+c_u  \jnl_u} \right)\,.
\label{ssAmp}
\end{multline}
Again, locality restricts to $X\ge1$.
 
  With these two building blocks, $\ctrivXY$ and $\cssXY$, we can construct any higher-derivative four-point adjoint-type amplitude that involves only adjoint color and scalar kinematics. Such amplitudes $s t \AHD(s,t)$ must be completely permutation invariant and thus can always be written in terms of a crossing-symmetric adjoint-type polynomial function of Mandelstam invariants $j(a,b,c)$ as:
\be
s t A^{\rm HD}(s,t) = c_s j_s + c_t j_t + c_u j_u\, .
\ee
Any such $j_s=j(s,t,u)$ can be written as a superposition of simple-scalar and NLSM numerators using the following general decomposition,
\be
j_s= \jss_s\left( \frac{\frac{t j_s}{t-u}+\frac{s j_t}{u-s}}{s-t} \right)
+\jnl_s \left( \frac{\frac{j_t}{s-u}-\frac{j_s}{t-u}}{s-t}\right)\,,
\label{generalJ}
\ee
a fact easily verified by recalling the definitions of $\jss(s,t,u)=(u-t)$ and $\jnl(s,t,u)=s(u-t)$. What is particularly notable is that their coefficients in \eqn{generalJ} are each invariant under all permutations $S_3(s,t,u)$ by virtue of the adjoint-type properties of $j(a,b,c)$. One might be concerned about potential poles, but both must in fact be local expressions. The simplest argument is to realize $b=c$ is always a zero of the polynomial $j(a,b,c)$ by virtue of antisymmetry, and thus $(b-c)$ must be a factor of $j(a,b,c)$. Similarly, as $s=t$ is manifestly a zero of each numerator in these expressions, the remaining divisor $(s-t)$ must be a factor of both. 

We have not yet exhausted all potential single-trace color modifications.  Namely, we have not yet considered the possibility that the color-weight information may itself be permutation invariant, as per the symmetric symbol: $\dabcd= \frac{1}{3!} \sum_{\sigma \in S_3(b,c,d)} \rm{Tr}(T^{a} T^{\sigma_1}T^{\sigma_2}T^{\sigma_3})\,.$ To make an adjoint-type building block, we simply take the product of this symbol with any of our two adjoint-type scalar numerators; due to redundancy between such products, we need only consider adding to our repertoire the final building block,
\be
\boxed{\cpermXY_s = \dabcd \jnl_s \pcub^X \psq^Y \ap^{(2+3X+2Y)}}\,.
\label{cNLSM}
\ee
These building blocks result in $(s,t)$ ordered amplitudes as,
\be
s t \permAXY(s,t)  =   \pcub^{X+1} \psq^Y \dabcd \,,
\label{dAmps}
\ee
again manifestly satisfying the usual field-theory relations by construction. 

With only three building blocks: $\ctrivXY$, $\cssXY$, and $\cpermXY$  we have exhausted all four-point single-trace higher-derivative modifications of color-weight, and so we find the generic form of such corrections to Yang-Mills to be encapsulated by:
\be
\label{gaugeSoln}
 \hat{c}_s  = \text{$\sum_{i}$} {\alpha'}^i (a_{XY} \ctrivXY_s +  a^{ss}_{XY} \cssXY_s+a^{\rm d}_{XY}  \cpermXY_s )
\ee
where the sum over  $X,Y$ relevant to mass dimension $\alpha'^i$ is left implicit, and the $a$ parameters encode distinct operator Wilson-coefficients.   See \Tab{operatorTable} for example scalar and YM operators through mass dimension four. As the unmodified Yang-Mills weights ensure compatibility with global SUSY, all local higher-derivative open supersstring corrections to the four point tree-level amplitude, consistent with adjoint-type representations, will be given by such $\hat{c}$ simply as:
\be
{\cal A}^{{\rm YM}+{\rm HD}}_4 =  \frac{\hat{c}_s \nym_s}{s}+\frac{\hat{c}_t \nym_t}{t}+\frac{\hat{c}_u \nym_u}{u}\,.
\label{fullHDYM}
\ee
We will see shortly that our simple color-modified building blocks for higher-derivative amplitudes are sufficient to capture the open superstring low-energy expansion~\cite{Bilal:2001hb, Broedel2012rc, Mafra2016mcc}.

We have, thus far, only modified color-weights. Composition between the above scalar weights and kinematic Yang-Mills weights always satisfies Jacobi identities, but it is easy to see that the only composition with our scalar weights that maintains gauge invariance is redundant with modifications to color we have already considered in \Eqn{trivialColor}.  What about non-superstring operators that can be applied to gauge theory? The same discussion carries through, essentially unchanged, by replacing $\nym$ with other  four-point adjoint color-dual vector weights $n^{\rm vec}$ not compositionally related to $\nym$ (cf.~Ref.\cite{Bern:2017tuc} for all seven distinct tensor structures) such as the $n^{\rm F^3}$  numerator weights identified in  Ref.~\cite{Broedel2012rc}. We will return to this point shortly  where we make it clear that only four additional vector building blocks are required to reproduce the open bosonic string. 

Next, let us consider higher-derivative gravitational operators.  We build corrections to \eqn{adjointGR} by replacing either $\tnym$ or $\nym$ or both with higher-derivative modified vector weights.  For greater than half-maximal local supersymmetry, involving building blocks that contribute to the open superstring means $\nym$ modified only trivially by scalar permutation invariants.  One can further consider replacing one or both vector numerators with the type of non-YM vector weights $n^{\rm vec}$ discussed in the previous paragraph. Such an example would be the dressings associated with a single insertion of  $F^3$ operators, $n^{\rm F^3}$, which contributes to the bosonic open string, and whose double-copy to gravity was considered in~\cite{Broedel2012rc,He:2016iqi}.  These are of particular interest because of the possibility of removing anomalies in associated supergravity theories~\cite{CarrascoN4Anomaly, BPRAnomalyCancel, Bern:2019isl}.  Indeed the counterterm considered in these papers is  given by one vector copy in ${\cal N}=4$ super Yang-Mills and one copy $n^{\rm F^3}$.  Including the remaining $n^{\rm vec}$ building blocks not appearing in the tree-level open string amplitudes broadens the scope for supersymmetric counterterms. For example, this allows for double-copy construction of the ${\cal N}=5$ supergravity counterterm of ref.~\cite{Bossard:2011tq} whose coefficient was found to be zero via explicit four-loop calculation in ref.~\cite{Bern:2014sna}. 

\section{String amplitudes at four points.} We are now prepared to discover our building blocks, resummed over all orders in $\alpha'$, in the tree-level four-point open supersymmetric string amplitude. This can be interpreted as answering a field theory question of how atoms of field-theory prediction  can be made consistent with a string-like UV completion. 
 We start by recognizing the open superstring amplitude as the field theory double copy between Chan-Paton dressed Z-theory~\cite{Broedel2013tta,Carrasco2016ldy} and supersymmetric Yang-Mills.  The four-point amplitude can be represented in a permutation-invariant color-dual form as:
\begin{align}
-{\cal A}^{\textrm{OSS}}=\frac{  [ s t A^{Z}(s,t) ]  \, [ s t \Aym(s,t) ]}{s t u}\, ,
\label{OSS}
\end{align}
where all supersymmetric Ward identities are satisfied by virtue of operations on the Yang-Mills factor $s t \Aym(s,t)$, and $A^{Z}(s,t)$ is the scalar field-theoretic ($s$-$t$) partial amplitude of Chan-Paton dressed Z-theory encoding all orders of $\ap$ corrections. We can build $[ s t A^{Z}(s,t) ]$ starting from the bi-ordered doubly-stripped partial Z-amplitude $Z_{1234}(s,u)$, where the subscript refers to the Chan-Paton trace ordering, and the parenthetical ordering obeys field-theory relations,
\be
Z_{1234}(s,u)=\frac{\ap^{-1}}{s \, u} \frac{\Gamma (1+ {\alpha'}\, s) \Gamma (1+ {\alpha'}\, t)}{ \Gamma ({\alpha'}\,(s+t))}
\ee
We form the field-theory permutation invariant for this Chan-Paton ordering by simply taking the product: $s\, u\, Z_{1234}(s,u) = s\, t\, Z_{1234}(s,t)$.  By exploiting monodromy relations to permute the subscript orderings \cite{Monodromy,Stieberger:2009hq}, we can generate the Chan-Paton dressed expression, required in \Eqn{OSS}, 
\begin{multline}
[ s t A^{Z}(s,t)]=  \!\!\!\! \sum_{\sigma\in{S_3(2,3,4)} }\!\!\!\! \text{Tr}_{[1\sigma]} s t Z_{1\sigma}(s,t)\\
 =\sum_{\sigma\in{S_3(2,3,4)} }\!\!\!\! \text{Tr}_{[1\sigma]} \frac{ \sin(\pi \ap s_{1,\sigma(3)})}{ \sin(\pi \ap s_{1,3})} \left( s t Z_{1234}(s,t) \right )\,,
 \end{multline}  
 where Tr${}_{[\rho]}$ denotes $\rm{Tr}[T^{a_{\rho(1)}}T^{a_{\rho(2)}}T^{a_{\rho(3)}}T^{a_{\rho(4)}}]$.
 The above is invariant under exchange of any channels, and by expressing the Chan-Paton trace factors in terms of $c_s$, $c_t$, $c_u$, and $\dabcd$, we find the following simple color-dual block-by-block form for Chan-Paton dressed Z-theory,
\be
\boxed{  \left [s t A^{Z}(s,t) \right]=\Gamma_{\{s,t,u\}} (\Zadj + \dabcd \Zsym) }\,
\ee
where $\Gamma_{\{s,t,u\}}$ corresponds to a series of higher mass-dimension combinations of Mandelstam invariants with coefficients responsible for familiar $\zeta$ contributions to the low-energy expansion. The factor   
$\Zadj$ contains all expressions involving Chan-Paton trace combinations $c_s$, $c_t$, or $c_u$, and $\Zsym$ contains all terms proportional to $\dabcd$.  These are given as follows:
\begin{align}
\Gamma_{\{s,t,u\}} &= \frac{\pi^2}{\ap} \frac{\csc(\pi \ap s) \csc(\pi \ap t) \csc(\pi \ap u)}{\Gamma({-\ap s}) \Gamma({-\ap t}) \Gamma({-\ap u})}\\ 
\Zsym &= 2 \, \left [ \sin({ \pi \ap s})+ \sin({ \pi \ap t})+ \sin({ \pi \ap u}) \right]\\
\Zadj &= c_s z_s + c_t z_t + c_u z_u
\end{align}
The $z_g$, which  satisfy anti-symmetry and Jacobi-identities to ensure the permutation invariance of $\Zadj$, take a particularly simple form, with  
\be
z_s = z(s,t,u)= (\sin( \pi \ap u) - \sin(\pi \ap t))/3\,,
\ee
In this form, all the coefficients for $\cpermXY$ may be easily identified already from the low-energy expansion of $\Zsym$. The remaining two building blocks only require a little teasing out from $\Zadj$. Introducing $S_p$ to denote $\sin(\pi \ap p)$, we can use \eqn{generalJ} to  express $z_g=  \jss_g \Zbi + \jnl_g \Zss$, with permutation invariant $\Zbi$ and $\Zss$ compactly given by:
 \begin{align}
 \Zbi&=\frac{1}{3} \frac{s (t-u) S_s +t (u-s) S_t +u (s-t) S_u}{(s - t) (s - u) (t - u) }\\
 \Zss&= \frac{1}{3}\frac{ (u-t) S_s +(s-u) S_t+(t-s) S_u}{(s - t) (s - u) (t - u) } \,.
 \end{align}

The $\Zbi$ terms in $\Zadj$ can thus be built from $\ctrivXY$,  and similarly the $\Zss$ terms from $\cssXY$. We have  exposed within the four-point open superstring amplitude the three unique Jacobi-identity satisfying modifications to the color-weights of Yang-Mills. The Z-theory block can now be written in terms of  
\begin{multline}
[s t A^{Z}(s,t)]=\stAbiAdj  \Gamma_{\{s,t,u\}} Z^{\rm bi}+\\
s t \AcssOO(s,t) \Gamma_{\{s,t,u\}} \Zss+d^{a_1 a_2 a_3 a_4} \Gamma_{\{s,t,u\}} \Zsym \, ,
\label{CPdressedZ}
\end{multline}
where the higher-derivative expressions through any mass dimension can be given by simply series expanding about $\ap\to0$. The individual $a_{XY}$  coefficients that fix \Eqn{fullHDYM} to the low-energy expansion of the open superstring through mass-dimension sixteen are given in an ancillary Mathematica file~\footnote{See the ancillary file of the arXiv version of this manuscript.}.

We now turn to the open bosonic string amplitude at four-point tree-level.  It was shown in Refs.~\cite{Huang:2016tag,Azevedo2018dgo} that this amplitude also obeys a field theoretic adjoint-type double-copy description with Z amplitudes as follows:
\begin{align}
\textrm{ (open bosonic) }=(\text {Z-theory}) \otimes\left(\textrm{YM}+(DF)^2\right)\, ,
\end{align}
where $(DF)^2$ is a massive higher-derivative YM theory, compatible with the usual BCJ relations but in violation of supersymmetric Ward identities. Only four new (distinct tensor~\cite{Bern:2017tuc}) building blocks are needed in the vector copy, along with permutation-invariant objects $\sigma_2$ and $\sigma_3$ compactly encoded in the following denominator:
\begin{multline}
A^{\textrm{YM}+(DF)^2}_4-A^{\textrm{YM}}_4 =\\
\frac{\ap A^{F^3}_4  +\ap^2 A^{(F^3)^2+F^4}_4+ \ap^3 A^{D^2F^4}_4+\ap^4 A^{ (DF)^4}}{(1-\alpha' s)(1-\alpha't)(1-\alpha' u)}\, .
\end{multline}
Explicit expressions for the four additional vector building blocks are provided in the auxiliary Mathematica file.

\section{Discussion.} We have found that a concrete understanding of all-order higher derivative corrections to YM and GR consistent with adjoint-type double-copy structure at four-points follows from a few field theory considerations. These corrections can be obtained through a simple composition rule that combines color-dual numerators into more complex numerators with the same algebraic properties, promoting the color-weights to carry higher-derivative corrections. Introducing three modified color-weights is sufficient to generate the predictions of every operator contributing to the low energy expansion of the tree-level open superstring, and we find that just four additional gauge-invariant building blocks, dressed with the same modified color factors, are sufficient for the tree-level bosonic string.  It remains an important open question as to whether all local supergravity operators at four-points have an adjoint-type double-copy structure. This seems decidable as we have shown that the number of viable  building blocks one needs to consider is quite small.

 These considerations only specify the analytic form of higher-derivative corrections. One may choose to fix their coefficients by assuming the asymptotic uniqueness of the Veneziano amplitude (cf.~Ref.~\cite{Barreiro:2012aw,Barreiro:2013dpa,Caron-Huot:2016icg}). This discussion complements and explains results noted in Ref. \cite{Carrasco:2019qwr}, demonstrating explicitly that the low-energy effective actions of super and bosonic strings, governed by Z-theory, are highly constrained by the duality between color and kinematics.

Preliminary exploration confirms~\cite{hdHigherMult} that the pattern of identifying color-dual building blocks that admit composition continues at higher-multiplicity, a topic that merits detailed study. Gaining all-multiplicity control will mean that, through unitarity methods, one could build relatively easy to construct higher loop-order scalar integrands that, via double copy, trivially recycle known gauge and gravity integrands to their higher-derivative corrections. 

It is worth remarking on the color-kinematic structure of the SUSY-compatible $F^4$ amplitude:
\be
{\cal A}^{F^4_{\rm SUSY} }= d^{a_1 a_2 a_3 a_4} \, s \,t\, \Aym(s,t) \, .
\ee
It was observed~\cite{Broedel2012rc} that the kinematic factor accompanying individual trace terms, $s\, t\, \Aym(s,t)$, does not satisfy the $(n-3)!$ relations associated with {\em adjoint} color-kinematic structure. While it is possible to misconstrue this result to show that $F^4$ is incompatible with color-kinematics duality in some broad sense, there are two striations of ${\cal A}^{F^4_{\rm SUSY} }$ along which we may see color-kinematics duality at play. 

First, from a perspective informed by many examples~\cite{Bargheer2012gv,Johansson2014zca,Johansson:2015oia,Johansson:2019dnu} of non-adjoint color-kinematics duality satisfying representations, we should emphasize that $F^4$ manifests a completely-symmetric color-kinematics duality: both the color term, $\dabcd$, and the kinematic (Born-Infeld) term,  $s t \Aym(s,t)$, are invariant under all permutations.  This seemingly trivial duality even has teeth: there is an associated double-copy construction. Replacing $\dabcd$ with the permutation-invariant kinematic weight $s t \Aym(s,t)$ generates the gravitational $R^4$ amplitude consistent with maximal local supersymmetry: $
{\cal A}^{R^4_{\rm SUSY} }=  (s \, t\, \Aym(s,t))^2 = s \, t \, u \, ({\cal A}^{\rm GR})$.
 
Second, we learn from \Eqns{cNLSM}{dAmps} that both ${\cal A}^{F^4}$ and ${\cal A}^{R^4}$ {\em also} manifest a non-trivial adjoint double-copy structure at four-points:
 ${\cal A}^{F^4_{\rm SUSY}}_4 = {\cal A}^{\rm YM}_4 |_{c_g\to (\dabcd \jnl_g)}$, and ${\cal A}^{R^4_{\rm SUSY}}_4 ={\cal A}^{\rm YM}_4 |_{c_g\to (\sigma_3  \nym_g)}$.
The key to realizing this adjoint-type color-dual representation is to allow both color factors and scalar kinematics to conspire within the same weight to satisfy the adjoint algebraic relations---a lesson driven home top-down by abelian Z-theory~\cite{Carrasco2016ldy}, and constructively presented here.

Not all effective particles are massless, and not all such particles are single-trace in the adjoint (cf. QCD with fermions in the fundamental, Einstein-Yang-Mills, and the standard model more generally), yet many admit color-dual representations~\cite{Johansson2014zca, Johansson:2015oia,Chiodaroli2017ngp,Du2017gnh, Johansson:2019dnu}. It will be fascinating to see how  constructive building blocks can encode  higher-derivative corrections to their predictions.  Generalizations of the above building blocks should be relevant to exploring higher-derivative corrections to  phenomenological effective field theories~\cite{Elvang:2018dco, Low:2019ynd, Carrillo-Gonzalez:2019aao,Bern:2019wie}.

\section{Acknowledgements}
We thank Zvi Bern, Marco Chiodaroli, Lance Dixon, André de Gouvêa, Henrik Johansson, Ian Low, Radu Roiban, Oliver Schlotterer, and Zhewei Yin for a combination of helpful discussions and comments on the manuscript. JJMC and LR are supported by the European Research Council under ERC-STG-639729, Strategic Predictions for Quantum Field Theories.

\begin{table*}
\begin{center}
\caption{Scalar and gauge operators corresponding to $\hat{c}$ through $\ap^4$.  For the bi-colored scalar operators we suppress their second color-indices and color-factors $\tilde{c}_s$. 
The  symmetrized trace operators generate the $\dabcd$ symbol as per~\cite{Bilal:2001hb}.}
\label{operatorTable}
\begin{tabular}{ c|c|c|c} 
\toprule
{\small Mass Dim.} & $\hat{c}_g $ & Scalar operator/$\tilde{c}_s$ & Gauge operator~\cite{Bilal:2001hb} \\
\midrule
2 & $c^{(0,0,d)}_g$ & $\dabcd(\partial_{\mu}\varphi_{a}\varphi_{b}\partial^{\mu}\varphi_{c}\varphi_{d})  $& $\str\left( 
F_{\mu\nu}F^{\nu\rho}F_{\rho \sigma}F^{\sigma \mu} 
-{\frac{1}{4}} \left( F_{\mu\nu}F^{\mu\nu}\right)^2 \right)$ \\
\midrule
3 & $c^{(1,0)}_g$ & $ \overbrace{f_{dae} f_{ebc} }^{c_t}(\partial_{\mu}\partial_{\nu}\varphi_{a}\partial^{\nu}\varphi_{b}\partial^{\mu}\varphi_{c}\varphi_{d}) $&$ c_t
\left\{ F^a_{\mu\nu} D_\lambda F^{b\nu}_{\ \ \rho}
( F^{c\rho}_{\ \ \sigma} D_\lambda F^{d\sigma \mu} 
+ F^{c\mu}_{\ \ \sigma} D_\lambda F^{d\sigma \rho} \right) $
 \\
&~& ~&$
-{ \frac{1}{2}} F^a_{\mu\nu} D_\lambda F^{b\mu\nu} F^c_{\rho \sigma} D_\lambda F^{d\rho \sigma}\}$ \\ 
\midrule
4 & $c^{(0,1,d)}_g$ & $\dabcd(\partial_{\mu}\partial_{\nu}\partial_{\rho}\varphi_{a}\partial^{\nu}\partial^{\rho}\varphi_{b}\partial^{\mu}\varphi_{c}\varphi_{d} ) $&$ \str \Big[ 
F_{\mu\nu} D^\lambda D^\kappa F^{\nu\rho} D_\lambda F_{\rho \sigma} D_\kappa F^{\sigma \mu} $
 \\
&~& ~&$
+{ \frac{1}{4}} F_{\mu\nu} D^\lambda D^\kappa F^{\mu\nu} D_\lambda F_{\rho \sigma} D_\kappa F^{\rho\sigma} \Big]$ \\
&~& ~& \\
\midrule
4 & $c^{(1,0,ss)}_g$ & $ ( \overbrace{f_{abe}f_{ecd}}^{c_s}+\overbrace{f_{cae}f_{edb}}^{c_u}) \times   $&$(c_{s}+c_{u}) \times $
 \\
&~& $(\partial_{\mu}\partial_{\nu}\partial_{\rho}\varphi_{a}\partial^{\rho}\varphi_{b}\partial^{\mu}\partial^{\nu}\varphi_{c}\varphi_{d}) $&$
\Big[ F^a_{\mu\nu} D^\lambda D^\kappa F^{b\nu\rho} D_\lambda F^c_{\rho \sigma} D_\kappa F^{d\sigma \mu} $ \\
&~&~& 
$ -{\frac{1}{4}}F^a_{\mu\nu} D^\lambda D^\kappa F^{b\mu\nu} D_\lambda F^c_{\rho \sigma} D_\kappa F^{d\rho \sigma} \Big] $ \\
&~& \\
\bottomrule
\end{tabular}
\end{center}
\end{table*}

\bibliography{simpleHD}

\begin{thebibliography}{47}%
\makeatletter
\providecommand \@ifxundefined [1]{%
 \@ifx{#1\undefined}
}%
\providecommand \@ifnum [1]{%
 \ifnum #1\expandafter \@firstoftwo
 \else \expandafter \@secondoftwo
 \fi
}%
\providecommand \@ifx [1]{%
 \ifx #1\expandafter \@firstoftwo
 \else \expandafter \@secondoftwo
 \fi
}%
\providecommand \natexlab [1]{#1}%
\providecommand \enquote  [1]{``#1''}%
\providecommand \bibnamefont  [1]{#1}%
\providecommand \bibfnamefont [1]{#1}%
\providecommand \citenamefont [1]{#1}%
\providecommand \href@noop [0]{\@secondoftwo}%
\providecommand \href [0]{\begingroup \@sanitize@url \@href}%
\providecommand \@href[1]{\@@startlink{#1}\@@href}%
\providecommand \@@href[1]{\endgroup#1\@@endlink}%
\providecommand \@sanitize@url [0]{\catcode `\\12\catcode `\$12\catcode
  `\&12\catcode `\#12\catcode `\^12\catcode `\_12\catcode `\%12\relax}%
\providecommand \@@startlink[1]{}%
\providecommand \@@endlink[0]{}%
\providecommand \url  [0]{\begingroup\@sanitize@url \@url }%
\providecommand \@url [1]{\endgroup\@href {#1}{\urlprefix }}%
\providecommand \urlprefix  [0]{URL }%
\providecommand \Eprint [0]{\href }%
\providecommand \doibase [0]{http://dx.doi.org/}%
\providecommand \selectlanguage [0]{\@gobble}%
\providecommand \bibinfo  [0]{\@secondoftwo}%
\providecommand \bibfield  [0]{\@secondoftwo}%
\providecommand \translation [1]{[#1]}%
\providecommand \BibitemOpen [0]{}%
\providecommand \bibitemStop [0]{}%
\providecommand \bibitemNoStop [0]{.\EOS\space}%
\providecommand \EOS [0]{\spacefactor3000\relax}%
\providecommand \BibitemShut  [1]{\csname bibitem#1\endcsname}%
\let\auto@bib@innerbib\@empty
\bibitem [{\citenamefont {Kawai}\ \emph {et~al.}(1986)\citenamefont {Kawai},
  \citenamefont {Lewellen},\ and\ \citenamefont {Tye}}]{KLT}%
  \BibitemOpen
  \bibfield  {author} {\bibinfo {author} {\bibfnamefont {H.}~\bibnamefont
  {Kawai}}, \bibinfo {author} {\bibfnamefont {D.~C.}\ \bibnamefont {Lewellen}},
  \ and\ \bibinfo {author} {\bibfnamefont {S.~H.~H.}\ \bibnamefont {Tye}},\
  }\bibfield  {title} {\enquote {\bibinfo {title} {{A relation between tree
  amplitudes of closed and open strings}},}\ }\href {\doibase
  10.1016/0550-3213(86)90362-7} {\bibfield  {journal} {\bibinfo  {journal}
  {Nucl. Phys.}\ }\textbf {\bibinfo {volume} {B269}},\ \bibinfo {pages} {1--23}
  (\bibinfo {year} {1986})}\BibitemShut {NoStop}%
\bibitem [{\citenamefont {Bern}\ \emph {et~al.}(2008)\citenamefont {Bern},
  \citenamefont {Carrasco},\ and\ \citenamefont {Johansson}}]{BCJ}%
  \BibitemOpen
  \bibfield  {author} {\bibinfo {author} {\bibfnamefont {Z.}~\bibnamefont
  {Bern}}, \bibinfo {author} {\bibfnamefont {J.~J.~M.}\ \bibnamefont
  {Carrasco}}, \ and\ \bibinfo {author} {\bibfnamefont {Henrik}\ \bibnamefont
  {Johansson}},\ }\bibfield  {title} {\enquote {\bibinfo {title} {{New
  relations for gauge-theory amplitudes}},}\ }\href {\doibase
  10.1103/PhysRevD.78.085011} {\bibfield  {journal} {\bibinfo  {journal} {Phys.
  Rev.}\ }\textbf {\bibinfo {volume} {D78}},\ \bibinfo {pages} {085011}
  (\bibinfo {year} {2008})},\ \Eprint {http://arxiv.org/abs/0805.3993}
  {arXiv:0805.3993 [hep-ph]} \BibitemShut {NoStop}%
\bibitem [{\citenamefont {Bern}\ \emph {et~al.}(2010)\citenamefont {Bern},
  \citenamefont {Carrasco},\ and\ \citenamefont {Johansson}}]{BCJLoop}%
  \BibitemOpen
  \bibfield  {author} {\bibinfo {author} {\bibfnamefont {Zvi}\ \bibnamefont
  {Bern}}, \bibinfo {author} {\bibfnamefont {John Joseph~M.}\ \bibnamefont
  {Carrasco}}, \ and\ \bibinfo {author} {\bibfnamefont {Henrik}\ \bibnamefont
  {Johansson}},\ }\bibfield  {title} {\enquote {\bibinfo {title} {{Perturbative
  quantum gravity as a double copy of gauge theory}},}\ }\href {\doibase
  10.1103/PhysRevLett.105.061602} {\bibfield  {journal} {\bibinfo  {journal}
  {Phys. Rev. Lett.}\ }\textbf {\bibinfo {volume} {105}},\ \bibinfo {pages}
  {061602} (\bibinfo {year} {2010})},\ \Eprint {http://arxiv.org/abs/1004.0476}
  {arXiv:1004.0476 [hep-th]} \BibitemShut {NoStop}%
\bibitem [{\citenamefont {Bern}\ \emph
  {et~al.}(2019{\natexlab{a}})\citenamefont {Bern}, \citenamefont {Carrasco},
  \citenamefont {Chiodaroli}, \citenamefont {Johansson},\ and\ \citenamefont
  {Roiban}}]{BCJreview}%
  \BibitemOpen
  \bibfield  {author} {\bibinfo {author} {\bibfnamefont {Zvi}\ \bibnamefont
  {Bern}}, \bibinfo {author} {\bibfnamefont {John~Joseph}\ \bibnamefont
  {Carrasco}}, \bibinfo {author} {\bibfnamefont {Marco}\ \bibnamefont
  {Chiodaroli}}, \bibinfo {author} {\bibfnamefont {Henrik}\ \bibnamefont
  {Johansson}}, \ and\ \bibinfo {author} {\bibfnamefont {Radu}\ \bibnamefont
  {Roiban}},\ }\bibfield  {title} {\enquote {\bibinfo {title} {{The Duality
  Between Color and Kinematics and its Applications}},}\ }\href@noop {} {\
  (\bibinfo {year} {2019}{\natexlab{a}})},\ \Eprint
  {http://arxiv.org/abs/1909.01358} {arXiv:1909.01358 [hep-th]} \BibitemShut
  {NoStop}%
\bibitem [{\citenamefont {Bern}\ \emph
  {et~al.}(2012{\natexlab{a}})\citenamefont {Bern}, \citenamefont {Carrasco},
  \citenamefont {Dixon}, \citenamefont {Johansson},\ and\ \citenamefont
  {Roiban}}]{SimplifyingBCJ}%
  \BibitemOpen
  \bibfield  {author} {\bibinfo {author} {\bibfnamefont {Z.}~\bibnamefont
  {Bern}}, \bibinfo {author} {\bibfnamefont {J.~J.~M.}\ \bibnamefont
  {Carrasco}}, \bibinfo {author} {\bibfnamefont {L.~J.}\ \bibnamefont {Dixon}},
  \bibinfo {author} {\bibfnamefont {H.}~\bibnamefont {Johansson}}, \ and\
  \bibinfo {author} {\bibfnamefont {R.}~\bibnamefont {Roiban}},\ }\bibfield
  {title} {\enquote {\bibinfo {title} {{Simplifying multiloop integrands and
  ultraviolet divergences of gauge theory and gravity amplitudes}},}\ }\href
  {\doibase 10.1103/PhysRevD.85.105014} {\bibfield  {journal} {\bibinfo
  {journal} {Phys. Rev.}\ }\textbf {\bibinfo {volume} {D85}},\ \bibinfo {pages}
  {105014} (\bibinfo {year} {2012}{\natexlab{a}})},\ \Eprint
  {http://arxiv.org/abs/1201.5366} {arXiv:1201.5366 [hep-th]} \BibitemShut
  {NoStop}%
\bibitem [{\citenamefont {Bern}\ \emph {et~al.}(2011)\citenamefont {Bern},
  \citenamefont {Boucher-Veronneau},\ and\ \citenamefont
  {Johansson}}]{N46Sugra}%
  \BibitemOpen
  \bibfield  {author} {\bibinfo {author} {\bibfnamefont {Z.}~\bibnamefont
  {Bern}}, \bibinfo {author} {\bibfnamefont {C.}~\bibnamefont
  {Boucher-Veronneau}}, \ and\ \bibinfo {author} {\bibfnamefont
  {H.}~\bibnamefont {Johansson}},\ }\bibfield  {title} {\enquote {\bibinfo
  {title} {{${\cal N} = 4$ supergravity amplitudes from gauge theory at one
  loop}},}\ }\href {\doibase 10.1103/PhysRevD.84.105035} {\bibfield  {journal}
  {\bibinfo  {journal} {Phys. Rev.}\ }\textbf {\bibinfo {volume} {D84}},\
  \bibinfo {pages} {105035} (\bibinfo {year} {2011})},\ \Eprint
  {http://arxiv.org/abs/1107.1935} {arXiv:1107.1935 [hep-th]} \BibitemShut
  {NoStop}%
\bibitem [{\citenamefont {Boucher-Veronneau}\ and\ \citenamefont
  {Dixon}(2011)}]{N46Sugra2}%
  \BibitemOpen
  \bibfield  {author} {\bibinfo {author} {\bibfnamefont {C.}~\bibnamefont
  {Boucher-Veronneau}}\ and\ \bibinfo {author} {\bibfnamefont {L.~J.}\
  \bibnamefont {Dixon}},\ }\bibfield  {title} {\enquote {\bibinfo {title}
  {{${\cal N}\ge 4$ supergravity amplitudes from gauge theory at two Loops}},}\
  }\href {\doibase 10.1007/JHEP12(2011)046} {\bibfield  {journal} {\bibinfo
  {journal} {JHEP}\ }\textbf {\bibinfo {volume} {12}},\ \bibinfo {pages} {046}
  (\bibinfo {year} {2011})},\ \Eprint {http://arxiv.org/abs/1110.1132}
  {arXiv:1110.1132 [hep-th]} \BibitemShut {NoStop}%
\bibitem [{\citenamefont {Bern}\ \emph
  {et~al.}(2012{\natexlab{b}})\citenamefont {Bern}, \citenamefont {Davies},
  \citenamefont {Dennen},\ and\ \citenamefont {Huang}}]{Bern:2012cd}%
  \BibitemOpen
  \bibfield  {author} {\bibinfo {author} {\bibfnamefont {Zvi}\ \bibnamefont
  {Bern}}, \bibinfo {author} {\bibfnamefont {Scott}\ \bibnamefont {Davies}},
  \bibinfo {author} {\bibfnamefont {Tristan}\ \bibnamefont {Dennen}}, \ and\
  \bibinfo {author} {\bibfnamefont {Yu-tin}\ \bibnamefont {Huang}},\ }\bibfield
   {title} {\enquote {\bibinfo {title} {{Absence of three-loop four-point
  divergences in ${\cal N}=4$ supergravity}},}\ }\href {\doibase
  10.1103/PhysRevLett.108.201301} {\bibfield  {journal} {\bibinfo  {journal}
  {Phys. Rev. Lett.}\ }\textbf {\bibinfo {volume} {108}},\ \bibinfo {pages}
  {201301} (\bibinfo {year} {2012}{\natexlab{b}})},\ \Eprint
  {http://arxiv.org/abs/1202.3423} {arXiv:1202.3423 [hep-th]} \BibitemShut
  {NoStop}%
\bibitem [{\citenamefont {Bern}\ \emph
  {et~al.}(2012{\natexlab{c}})\citenamefont {Bern}, \citenamefont {Davies},
  \citenamefont {Dennen},\ and\ \citenamefont {Huang}}]{Bern:2012gh}%
  \BibitemOpen
  \bibfield  {author} {\bibinfo {author} {\bibfnamefont {Zvi}\ \bibnamefont
  {Bern}}, \bibinfo {author} {\bibfnamefont {Scott}\ \bibnamefont {Davies}},
  \bibinfo {author} {\bibfnamefont {Tristan}\ \bibnamefont {Dennen}}, \ and\
  \bibinfo {author} {\bibfnamefont {Yu-tin}\ \bibnamefont {Huang}},\ }\bibfield
   {title} {\enquote {\bibinfo {title} {{Ultraviolet cancellations in
  half-maximal supergravity as a consequence of the double-copy structure}},}\
  }\href {\doibase 10.1103/PhysRevD.86.105014} {\bibfield  {journal} {\bibinfo
  {journal} {Phys. Rev.}\ }\textbf {\bibinfo {volume} {D86}},\ \bibinfo {pages}
  {105014} (\bibinfo {year} {2012}{\natexlab{c}})},\ \Eprint
  {http://arxiv.org/abs/1209.2472} {arXiv:1209.2472 [hep-th]} \BibitemShut
  {NoStop}%
\bibitem [{\citenamefont {Boels}\ and\ \citenamefont
  {Isermann}(2013)}]{Boels:2012sy}%
  \BibitemOpen
  \bibfield  {author} {\bibinfo {author} {\bibfnamefont {Rutger~H.}\
  \bibnamefont {Boels}}\ and\ \bibinfo {author} {\bibfnamefont {Reinke~Sven}\
  \bibnamefont {Isermann}},\ }\bibfield  {title} {\enquote {\bibinfo {title}
  {{On powercounting in perturbative quantum gravity theories through
  color-kinematic duality}},}\ }\href {\doibase 10.1007/JHEP06(2013)017}
  {\bibfield  {journal} {\bibinfo  {journal} {JHEP}\ }\textbf {\bibinfo
  {volume} {06}},\ \bibinfo {pages} {017} (\bibinfo {year} {2013})},\ \Eprint
  {http://arxiv.org/abs/1212.3473} {arXiv:1212.3473 [hep-th]} \BibitemShut
  {NoStop}%
\bibitem [{\citenamefont {Bern}\ \emph {et~al.}(2013)\citenamefont {Bern},
  \citenamefont {Davies},\ and\ \citenamefont {Dennen}}]{Bern:2013qca}%
  \BibitemOpen
  \bibfield  {author} {\bibinfo {author} {\bibfnamefont {Zvi}\ \bibnamefont
  {Bern}}, \bibinfo {author} {\bibfnamefont {Scott}\ \bibnamefont {Davies}}, \
  and\ \bibinfo {author} {\bibfnamefont {Tristan}\ \bibnamefont {Dennen}},\
  }\bibfield  {title} {\enquote {\bibinfo {title} {{The ultraviolet structure
  of half-maximal supergravity with matter multiplets at two and three
  loops}},}\ }\href {\doibase 10.1103/PhysRevD.88.065007} {\bibfield  {journal}
  {\bibinfo  {journal} {Phys. Rev.}\ }\textbf {\bibinfo {volume} {D88}},\
  \bibinfo {pages} {065007} (\bibinfo {year} {2013})},\ \Eprint
  {http://arxiv.org/abs/1305.4876} {arXiv:1305.4876 [hep-th]} \BibitemShut
  {NoStop}%
\bibitem [{\citenamefont {Bern}\ \emph
  {et~al.}(2014{\natexlab{a}})\citenamefont {Bern}, \citenamefont {Davies},\
  and\ \citenamefont {Dennen}}]{Bern:2014lha}%
  \BibitemOpen
  \bibfield  {author} {\bibinfo {author} {\bibfnamefont {Zvi}\ \bibnamefont
  {Bern}}, \bibinfo {author} {\bibfnamefont {Scott}\ \bibnamefont {Davies}}, \
  and\ \bibinfo {author} {\bibfnamefont {Tristan}\ \bibnamefont {Dennen}},\
  }\bibfield  {title} {\enquote {\bibinfo {title} {{The ultraviolet critical
  dimension of half-maximal supergravity at three loops}},}\ }\href@noop {} {\
  (\bibinfo {year} {2014}{\natexlab{a}})},\ \Eprint
  {http://arxiv.org/abs/1412.2441} {arXiv:1412.2441 [hep-th]} \BibitemShut
  {NoStop}%
\bibitem [{\citenamefont {Bern}\ \emph
  {et~al.}(2018{\natexlab{a}})\citenamefont {Bern}, \citenamefont {Carrasco},
  \citenamefont {Chen}, \citenamefont {Edison}, \citenamefont {Johansson},
  \citenamefont {Parra-Martinez}, \citenamefont {Roiban},\ and\ \citenamefont
  {Zeng}}]{UVFiveLoops}%
  \BibitemOpen
  \bibfield  {author} {\bibinfo {author} {\bibfnamefont {Zvi}\ \bibnamefont
  {Bern}}, \bibinfo {author} {\bibfnamefont {John~Joseph}\ \bibnamefont
  {Carrasco}}, \bibinfo {author} {\bibfnamefont {Wei-Ming}\ \bibnamefont
  {Chen}}, \bibinfo {author} {\bibfnamefont {Alex}\ \bibnamefont {Edison}},
  \bibinfo {author} {\bibfnamefont {Henrik}\ \bibnamefont {Johansson}},
  \bibinfo {author} {\bibfnamefont {Julio}\ \bibnamefont {Parra-Martinez}},
  \bibinfo {author} {\bibfnamefont {Radu}\ \bibnamefont {Roiban}}, \ and\
  \bibinfo {author} {\bibfnamefont {Mao}\ \bibnamefont {Zeng}},\ }\bibfield
  {title} {\enquote {\bibinfo {title} {{Ultraviolet properties of $\mathcal N =
  8$ supergravity at five loops}},}\ }\href {\doibase
  10.1103/PhysRevD.98.086021} {\bibfield  {journal} {\bibinfo  {journal} {Phys.
  Rev.}\ }\textbf {\bibinfo {volume} {D98}},\ \bibinfo {pages} {086021}
  (\bibinfo {year} {2018}{\natexlab{a}})},\ \Eprint
  {http://arxiv.org/abs/1804.09311} {arXiv:1804.09311 [hep-th]} \BibitemShut
  {NoStop}%
\bibitem [{\citenamefont {Herrmann}\ and\ \citenamefont
  {Trnka}(2016)}]{Herrmann:2016qea}%
  \BibitemOpen
  \bibfield  {author} {\bibinfo {author} {\bibfnamefont {Enrico}\ \bibnamefont
  {Herrmann}}\ and\ \bibinfo {author} {\bibfnamefont {Jaroslav}\ \bibnamefont
  {Trnka}},\ }\bibfield  {title} {\enquote {\bibinfo {title} {{Gravity On-shell
  diagrams}},}\ }\href {\doibase 10.1007/JHEP11(2016)136} {\bibfield  {journal}
  {\bibinfo  {journal} {JHEP}\ }\textbf {\bibinfo {volume} {11}},\ \bibinfo
  {pages} {136} (\bibinfo {year} {2016})},\ \Eprint
  {http://arxiv.org/abs/1604.03479} {arXiv:1604.03479 [hep-th]} \BibitemShut
  {NoStop}%
\bibitem [{\citenamefont {Herrmann}\ and\ \citenamefont
  {Trnka}(2019)}]{HerrmannTrnkaUVGrav}%
  \BibitemOpen
  \bibfield  {author} {\bibinfo {author} {\bibfnamefont {Enrico}\ \bibnamefont
  {Herrmann}}\ and\ \bibinfo {author} {\bibfnamefont {Jaroslav}\ \bibnamefont
  {Trnka}},\ }\bibfield  {title} {\enquote {\bibinfo {title} {{UV cancellations
  in gravity loop integrands}},}\ }\href {\doibase 10.1007/JHEP02(2019)084}
  {\bibfield  {journal} {\bibinfo  {journal} {JHEP}\ }\textbf {\bibinfo
  {volume} {02}},\ \bibinfo {pages} {084} (\bibinfo {year} {2019})},\ \Eprint
  {http://arxiv.org/abs/1808.10446} {arXiv:1808.10446 [hep-th]} \BibitemShut
  {NoStop}%
\bibitem [{\citenamefont {Broedel}\ \emph {et~al.}(2013)\citenamefont
  {Broedel}, \citenamefont {Schlotterer},\ and\ \citenamefont
  {Stieberger}}]{Broedel2013tta}%
  \BibitemOpen
  \bibfield  {author} {\bibinfo {author} {\bibfnamefont {Johannes}\
  \bibnamefont {Broedel}}, \bibinfo {author} {\bibfnamefont {Oliver}\
  \bibnamefont {Schlotterer}}, \ and\ \bibinfo {author} {\bibfnamefont
  {Stephan}\ \bibnamefont {Stieberger}},\ }\bibfield  {title} {\enquote
  {\bibinfo {title} {{Polylogarithms, multiple zeta values and superstring
  amplitudes}},}\ }\href {\doibase 10.1002/prop.201300019} {\bibfield
  {journal} {\bibinfo  {journal} {Fortsch. Phys.}\ }\textbf {\bibinfo {volume}
  {61}},\ \bibinfo {pages} {812--870} (\bibinfo {year} {2013})},\ \Eprint
  {http://arxiv.org/abs/1304.7267} {arXiv:1304.7267 [hep-th]} \BibitemShut
  {NoStop}%
\bibitem [{\citenamefont {Huang}\ \emph {et~al.}(2016)\citenamefont {Huang},
  \citenamefont {Schlotterer},\ and\ \citenamefont {Wen}}]{Huang:2016tag}%
  \BibitemOpen
  \bibfield  {author} {\bibinfo {author} {\bibfnamefont {Yu-tin}\ \bibnamefont
  {Huang}}, \bibinfo {author} {\bibfnamefont {Oliver}\ \bibnamefont
  {Schlotterer}}, \ and\ \bibinfo {author} {\bibfnamefont {Congkao}\
  \bibnamefont {Wen}},\ }\bibfield  {title} {\enquote {\bibinfo {title}
  {{Universality in string interactions}},}\ }\href {\doibase
  10.1007/JHEP09(2016)155} {\bibfield  {journal} {\bibinfo  {journal} {JHEP}\
  }\textbf {\bibinfo {volume} {09}},\ \bibinfo {pages} {155} (\bibinfo {year}
  {2016})},\ \Eprint {http://arxiv.org/abs/1602.01674} {arXiv:1602.01674
  [hep-th]} \BibitemShut {NoStop}%
\bibitem [{\citenamefont {Carrasco}\ \emph
  {et~al.}(2017{\natexlab{a}})\citenamefont {Carrasco}, \citenamefont {Mafra},\
  and\ \citenamefont {Schlotterer}}]{Carrasco2016ldy}%
  \BibitemOpen
  \bibfield  {author} {\bibinfo {author} {\bibfnamefont {John Joseph~M.}\
  \bibnamefont {Carrasco}}, \bibinfo {author} {\bibfnamefont {Carlos~R.}\
  \bibnamefont {Mafra}}, \ and\ \bibinfo {author} {\bibfnamefont {Oliver}\
  \bibnamefont {Schlotterer}},\ }\bibfield  {title} {\enquote {\bibinfo {title}
  {{Abelian Z-theory: NLSM amplitudes and $\alpha$'-corrections from the open
  string}},}\ }\href {\doibase 10.1007/JHEP06(2017)093} {\bibfield  {journal}
  {\bibinfo  {journal} {JHEP}\ }\textbf {\bibinfo {volume} {06}},\ \bibinfo
  {pages} {093} (\bibinfo {year} {2017}{\natexlab{a}})},\ \Eprint
  {http://arxiv.org/abs/1608.02569} {arXiv:1608.02569 [hep-th]} \BibitemShut
  {NoStop}%
\bibitem [{\citenamefont {Mafra}\ and\ \citenamefont
  {Schlotterer}(2017)}]{Mafra2016mcc}%
  \BibitemOpen
  \bibfield  {author} {\bibinfo {author} {\bibfnamefont {Carlos~R.}\
  \bibnamefont {Mafra}}\ and\ \bibinfo {author} {\bibfnamefont {Oliver}\
  \bibnamefont {Schlotterer}},\ }\bibfield  {title} {\enquote {\bibinfo {title}
  {{Non-abelian $Z$-theory: Berends-Giele recursion for the $\alpha'$-expansion
  of disk integrals}},}\ }\href {\doibase 10.1007/JHEP01(2017)031} {\bibfield
  {journal} {\bibinfo  {journal} {JHEP}\ }\textbf {\bibinfo {volume} {01}},\
  \bibinfo {pages} {031} (\bibinfo {year} {2017})},\ \Eprint
  {http://arxiv.org/abs/1609.07078} {arXiv:1609.07078 [hep-th]} \BibitemShut
  {NoStop}%
\bibitem [{\citenamefont {Carrasco}\ \emph
  {et~al.}(2017{\natexlab{b}})\citenamefont {Carrasco}, \citenamefont {Mafra},\
  and\ \citenamefont {Schlotterer}}]{Carrasco2016ygv}%
  \BibitemOpen
  \bibfield  {author} {\bibinfo {author} {\bibfnamefont {John Joseph~M.}\
  \bibnamefont {Carrasco}}, \bibinfo {author} {\bibfnamefont {Carlos~R.}\
  \bibnamefont {Mafra}}, \ and\ \bibinfo {author} {\bibfnamefont {Oliver}\
  \bibnamefont {Schlotterer}},\ }\bibfield  {title} {\enquote {\bibinfo {title}
  {{Semi-abelian Z-theory: NLSM$+\phi^{3}$ from the open string}},}\ }\href
  {\doibase 10.1007/JHEP08(2017)135} {\bibfield  {journal} {\bibinfo  {journal}
  {JHEP}\ }\textbf {\bibinfo {volume} {08}},\ \bibinfo {pages} {135} (\bibinfo
  {year} {2017}{\natexlab{b}})},\ \Eprint {http://arxiv.org/abs/1612.06446}
  {arXiv:1612.06446 [hep-th]} \BibitemShut {NoStop}%
\bibitem [{\citenamefont {Azevedo}\ \emph {et~al.}(2018)\citenamefont
  {Azevedo}, \citenamefont {Chiodaroli}, \citenamefont {Johansson},\ and\
  \citenamefont {Schlotterer}}]{Azevedo2018dgo}%
  \BibitemOpen
  \bibfield  {author} {\bibinfo {author} {\bibfnamefont {Thales}\ \bibnamefont
  {Azevedo}}, \bibinfo {author} {\bibfnamefont {Marco}\ \bibnamefont
  {Chiodaroli}}, \bibinfo {author} {\bibfnamefont {Henrik}\ \bibnamefont
  {Johansson}}, \ and\ \bibinfo {author} {\bibfnamefont {Oliver}\ \bibnamefont
  {Schlotterer}},\ }\bibfield  {title} {\enquote {\bibinfo {title} {{Heterotic
  and bosonic string amplitudes via field theory}},}\ }\href {\doibase
  10.1007/JHEP10(2018)012} {\bibfield  {journal} {\bibinfo  {journal} {JHEP}\
  }\textbf {\bibinfo {volume} {10}},\ \bibinfo {pages} {012} (\bibinfo {year}
  {2018})},\ \Eprint {http://arxiv.org/abs/1803.05452} {arXiv:1803.05452
  [hep-th]} \BibitemShut {NoStop}%
\bibitem [{\citenamefont {Bern}\ \emph {et~al.}(2017)\citenamefont {Bern},
  \citenamefont {Edison}, \citenamefont {Kosower},\ and\ \citenamefont
  {Parra-Martinez}}]{Bern:2017tuc}%
  \BibitemOpen
  \bibfield  {author} {\bibinfo {author} {\bibfnamefont {Zvi}\ \bibnamefont
  {Bern}}, \bibinfo {author} {\bibfnamefont {Alex}\ \bibnamefont {Edison}},
  \bibinfo {author} {\bibfnamefont {David}\ \bibnamefont {Kosower}}, \ and\
  \bibinfo {author} {\bibfnamefont {Julio}\ \bibnamefont {Parra-Martinez}},\
  }\bibfield  {title} {\enquote {\bibinfo {title} {{Curvature-squared
  multiplets, evanescent effects, and the U(1) anomaly in ${\cal N}=4$
  supergravity}},}\ }\href {\doibase 10.1103/PhysRevD.96.066004} {\bibfield
  {journal} {\bibinfo  {journal} {Phys. Rev.}\ }\textbf {\bibinfo {volume}
  {D96}},\ \bibinfo {pages} {066004} (\bibinfo {year} {2017})},\ \Eprint
  {http://arxiv.org/abs/1706.01486} {arXiv:1706.01486 [hep-th]} \BibitemShut
  {NoStop}%
\bibitem [{\citenamefont {Bilal}(2001)}]{Bilal:2001hb}%
  \BibitemOpen
  \bibfield  {author} {\bibinfo {author} {\bibfnamefont {Adel}\ \bibnamefont
  {Bilal}},\ }\bibfield  {title} {\enquote {\bibinfo {title} {{Higher
  derivative corrections to the nonAbelian Born-Infeld action}},}\ }\href
  {\doibase 10.1016/S0550-3213(01)00472-2} {\bibfield  {journal} {\bibinfo
  {journal} {Nucl. Phys.}\ }\textbf {\bibinfo {volume} {B618}},\ \bibinfo
  {pages} {21--49} (\bibinfo {year} {2001})},\ \Eprint
  {http://arxiv.org/abs/hep-th/0106062} {arXiv:hep-th/0106062 [hep-th]}
  \BibitemShut {NoStop}%
\bibitem [{\citenamefont {Broedel}\ and\ \citenamefont
  {Dixon}(2012)}]{Broedel2012rc}%
  \BibitemOpen
  \bibfield  {author} {\bibinfo {author} {\bibfnamefont {Johannes}\
  \bibnamefont {Broedel}}\ and\ \bibinfo {author} {\bibfnamefont {Lance~J.}\
  \bibnamefont {Dixon}},\ }\bibfield  {title} {\enquote {\bibinfo {title}
  {{Color-kinematics duality and double-copy construction for amplitudes from
  higher-dimension operators}},}\ }\href {\doibase 10.1007/JHEP10(2012)091}
  {\bibfield  {journal} {\bibinfo  {journal} {JHEP}\ }\textbf {\bibinfo
  {volume} {10}},\ \bibinfo {pages} {091} (\bibinfo {year} {2012})},\ \Eprint
  {http://arxiv.org/abs/1208.0876} {arXiv:1208.0876 [hep-th]} \BibitemShut
  {NoStop}%
\bibitem [{\citenamefont {He}\ and\ \citenamefont {Zhang}(2017)}]{He:2016iqi}%
  \BibitemOpen
  \bibfield  {author} {\bibinfo {author} {\bibfnamefont {Song}\ \bibnamefont
  {He}}\ and\ \bibinfo {author} {\bibfnamefont {Yong}\ \bibnamefont {Zhang}},\
  }\bibfield  {title} {\enquote {\bibinfo {title} {{New formulas for amplitudes
  from higher-dimensional operators}},}\ }\href {\doibase
  10.1007/JHEP02(2017)019} {\bibfield  {journal} {\bibinfo  {journal} {JHEP}\
  }\textbf {\bibinfo {volume} {02}},\ \bibinfo {pages} {019} (\bibinfo {year}
  {2017})},\ \Eprint {http://arxiv.org/abs/1608.08448} {arXiv:1608.08448
  [hep-th]} \BibitemShut {NoStop}%
\bibitem [{\citenamefont {Carrasco}\ \emph {et~al.}(2013)\citenamefont
  {Carrasco}, \citenamefont {Kallosh}, \citenamefont {Roiban},\ and\
  \citenamefont {Tseytlin}}]{CarrascoN4Anomaly}%
  \BibitemOpen
  \bibfield  {author} {\bibinfo {author} {\bibfnamefont {J.~J.~M.}\
  \bibnamefont {Carrasco}}, \bibinfo {author} {\bibfnamefont {R.}~\bibnamefont
  {Kallosh}}, \bibinfo {author} {\bibfnamefont {R.}~\bibnamefont {Roiban}}, \
  and\ \bibinfo {author} {\bibfnamefont {A.~A.}\ \bibnamefont {Tseytlin}},\
  }\bibfield  {title} {\enquote {\bibinfo {title} {{On the U(1) duality anomaly
  and the S-matrix of ${\cal N}=4$ supergravity}},}\ }\href {\doibase
  10.1007/JHEP07(2013)029} {\bibfield  {journal} {\bibinfo  {journal} {JHEP}\
  }\textbf {\bibinfo {volume} {07}},\ \bibinfo {pages} {029} (\bibinfo {year}
  {2013})},\ \Eprint {http://arxiv.org/abs/1303.6219} {arXiv:1303.6219
  [hep-th]} \BibitemShut {NoStop}%
\bibitem [{\citenamefont {Bern}\ \emph
  {et~al.}(2018{\natexlab{b}})\citenamefont {Bern}, \citenamefont
  {Parra-Martinez},\ and\ \citenamefont {Roiban}}]{BPRAnomalyCancel}%
  \BibitemOpen
  \bibfield  {author} {\bibinfo {author} {\bibfnamefont {Zvi}\ \bibnamefont
  {Bern}}, \bibinfo {author} {\bibfnamefont {Julio}\ \bibnamefont
  {Parra-Martinez}}, \ and\ \bibinfo {author} {\bibfnamefont {Radu}\
  \bibnamefont {Roiban}},\ }\bibfield  {title} {\enquote {\bibinfo {title}
  {{Canceling the U(1) anomaly in the $S$ matrix of ${\cal N}=4$
  supergravity}},}\ }\href {\doibase 10.1103/PhysRevLett.121.101604} {\bibfield
   {journal} {\bibinfo  {journal} {Phys. Rev. Lett.}\ }\textbf {\bibinfo
  {volume} {121}},\ \bibinfo {pages} {101604} (\bibinfo {year}
  {2018}{\natexlab{b}})},\ \Eprint {http://arxiv.org/abs/1712.03928}
  {arXiv:1712.03928 [hep-th]} \BibitemShut {NoStop}%
\bibitem [{\citenamefont {Bern}\ \emph
  {et~al.}(2019{\natexlab{b}})\citenamefont {Bern}, \citenamefont {Kosower},\
  and\ \citenamefont {Parra-Martinez}}]{Bern:2019isl}%
  \BibitemOpen
  \bibfield  {author} {\bibinfo {author} {\bibfnamefont {Zvi}\ \bibnamefont
  {Bern}}, \bibinfo {author} {\bibfnamefont {David}\ \bibnamefont {Kosower}}, \
  and\ \bibinfo {author} {\bibfnamefont {Julio}\ \bibnamefont
  {Parra-Martinez}},\ }\bibfield  {title} {\enquote {\bibinfo {title}
  {{Two-loop $n$-point anomalous amplitudes in ${\cal N}=4$ supergravity}},}\
  }\href@noop {} {\  (\bibinfo {year} {2019}{\natexlab{b}})},\ \Eprint
  {http://arxiv.org/abs/1905.05151} {arXiv:1905.05151 [hep-th]} \BibitemShut
  {NoStop}%
\bibitem [{\citenamefont {Bossard}\ \emph {et~al.}(2011)\citenamefont
  {Bossard}, \citenamefont {Howe}, \citenamefont {Stelle},\ and\ \citenamefont
  {Vanhove}}]{Bossard:2011tq}%
  \BibitemOpen
  \bibfield  {author} {\bibinfo {author} {\bibfnamefont {Guillaume}\
  \bibnamefont {Bossard}}, \bibinfo {author} {\bibfnamefont {P.~S.}\
  \bibnamefont {Howe}}, \bibinfo {author} {\bibfnamefont {K.~S.}\ \bibnamefont
  {Stelle}}, \ and\ \bibinfo {author} {\bibfnamefont {Pierre}\ \bibnamefont
  {Vanhove}},\ }\bibfield  {title} {\enquote {\bibinfo {title} {{The vanishing
  volume of D=4 superspace}},}\ }\href {\doibase
  10.1088/0264-9381/28/21/215005} {\bibfield  {journal} {\bibinfo  {journal}
  {Class. Quant. Grav.}\ }\textbf {\bibinfo {volume} {28}},\ \bibinfo {pages}
  {215005} (\bibinfo {year} {2011})},\ \Eprint {http://arxiv.org/abs/1105.6087}
  {arXiv:1105.6087 [hep-th]} \BibitemShut {NoStop}%
\bibitem [{\citenamefont {Bern}\ \emph
  {et~al.}(2014{\natexlab{b}})\citenamefont {Bern}, \citenamefont {Davies},\
  and\ \citenamefont {Dennen}}]{Bern:2014sna}%
  \BibitemOpen
  \bibfield  {author} {\bibinfo {author} {\bibfnamefont {Zvi}\ \bibnamefont
  {Bern}}, \bibinfo {author} {\bibfnamefont {Scott}\ \bibnamefont {Davies}}, \
  and\ \bibinfo {author} {\bibfnamefont {Tristan}\ \bibnamefont {Dennen}},\
  }\bibfield  {title} {\enquote {\bibinfo {title} {{Enhanced ultraviolet
  cancellations in $\mathcal N=5$ supergravity at four loops}},}\ }\href
  {\doibase 10.1103/PhysRevD.90.105011} {\bibfield  {journal} {\bibinfo
  {journal} {Phys. Rev. D}\ }\textbf {\bibinfo {volume} {90}},\ \bibinfo
  {pages} {105011} (\bibinfo {year} {2014}{\natexlab{b}})},\ \Eprint
  {http://arxiv.org/abs/1409.3089} {arXiv:1409.3089 [hep-th]} \BibitemShut
  {NoStop}%
\bibitem [{\citenamefont {Bjerrum-Bohr}\ \emph {et~al.}(2009)\citenamefont
  {Bjerrum-Bohr}, \citenamefont {Damgaard},\ and\ \citenamefont
  {Vanhove}}]{Monodromy}%
  \BibitemOpen
  \bibfield  {author} {\bibinfo {author} {\bibfnamefont {N.~E.~J.}\
  \bibnamefont {Bjerrum-Bohr}}, \bibinfo {author} {\bibfnamefont {Poul~H.}\
  \bibnamefont {Damgaard}}, \ and\ \bibinfo {author} {\bibfnamefont {Pierre}\
  \bibnamefont {Vanhove}},\ }\bibfield  {title} {\enquote {\bibinfo {title}
  {{Minimal basis for gauge theory amplitudes}},}\ }\href {\doibase
  10.1103/PhysRevLett.103.161602} {\bibfield  {journal} {\bibinfo  {journal}
  {Phys. Rev. Lett.}\ }\textbf {\bibinfo {volume} {103}},\ \bibinfo {pages}
  {161602} (\bibinfo {year} {2009})},\ \Eprint {http://arxiv.org/abs/0907.1425}
  {arXiv:0907.1425 [hep-th]} \BibitemShut {NoStop}%
\bibitem [{\citenamefont {Stieberger}(2009)}]{Stieberger:2009hq}%
  \BibitemOpen
  \bibfield  {author} {\bibinfo {author} {\bibfnamefont {S.}~\bibnamefont
  {Stieberger}},\ }\bibfield  {title} {\enquote {\bibinfo {title} {{Open and
  closed vs. pure open string Disk amplitudes}},}\ }\href@noop {} {\  (\bibinfo
  {year} {2009})},\ \Eprint {http://arxiv.org/abs/0907.2211} {arXiv:0907.2211
  [hep-th]} \BibitemShut {NoStop}%
\bibitem [{\citenamefont {Barreiro}\ and\ \citenamefont
  {Medina}(2012)}]{Barreiro:2012aw}%
  \BibitemOpen
  \bibfield  {author} {\bibinfo {author} {\bibfnamefont {Luiz~Antonio}\
  \bibnamefont {Barreiro}}\ and\ \bibinfo {author} {\bibfnamefont {Ricardo}\
  \bibnamefont {Medina}},\ }\bibfield  {title} {\enquote {\bibinfo {title}
  {{Revisiting the S-matrix approach to the open superstring low energy
  effective lagrangian}},}\ }\href {\doibase 10.1007/JHEP10(2012)108}
  {\bibfield  {journal} {\bibinfo  {journal} {JHEP}\ }\textbf {\bibinfo
  {volume} {10}},\ \bibinfo {pages} {108} (\bibinfo {year} {2012})},\ \Eprint
  {http://arxiv.org/abs/1208.6066} {arXiv:1208.6066 [hep-th]} \BibitemShut
  {NoStop}%
\bibitem [{\citenamefont {Barreiro}\ and\ \citenamefont
  {Medina}(2014)}]{Barreiro:2013dpa}%
  \BibitemOpen
  \bibfield  {author} {\bibinfo {author} {\bibfnamefont {Luiz~Antonio}\
  \bibnamefont {Barreiro}}\ and\ \bibinfo {author} {\bibfnamefont {Ricardo}\
  \bibnamefont {Medina}},\ }\bibfield  {title} {\enquote {\bibinfo {title}
  {{RNS derivation of $N$-point disk amplitudes from the revisited S-matrix
  approach}},}\ }\href {\doibase 10.1016/j.nuclphysb.2014.07.015} {\bibfield
  {journal} {\bibinfo  {journal} {Nucl. Phys.}\ }\textbf {\bibinfo {volume}
  {B886}},\ \bibinfo {pages} {870--951} (\bibinfo {year} {2014})},\ \Eprint
  {http://arxiv.org/abs/1310.5942} {arXiv:1310.5942 [hep-th]} \BibitemShut
  {NoStop}%
\bibitem [{\citenamefont {Caron-Huot}\ \emph {et~al.}(2017)\citenamefont
  {Caron-Huot}, \citenamefont {Komargodski}, \citenamefont {Sever},\ and\
  \citenamefont {Zhiboedov}}]{Caron-Huot:2016icg}%
  \BibitemOpen
  \bibfield  {author} {\bibinfo {author} {\bibfnamefont {Simon}\ \bibnamefont
  {Caron-Huot}}, \bibinfo {author} {\bibfnamefont {Zohar}\ \bibnamefont
  {Komargodski}}, \bibinfo {author} {\bibfnamefont {Amit}\ \bibnamefont
  {Sever}}, \ and\ \bibinfo {author} {\bibfnamefont {Alexander}\ \bibnamefont
  {Zhiboedov}},\ }\bibfield  {title} {\enquote {\bibinfo {title} {{Strings from
  Massive Higher Spins: The Asymptotic Uniqueness of the Veneziano
  Amplitude}},}\ }\href {\doibase 10.1007/JHEP10(2017)026} {\bibfield
  {journal} {\bibinfo  {journal} {JHEP}\ }\textbf {\bibinfo {volume} {10}},\
  \bibinfo {pages} {026} (\bibinfo {year} {2017})},\ \Eprint
  {http://arxiv.org/abs/1607.04253} {arXiv:1607.04253 [hep-th]} \BibitemShut
  {NoStop}%
\bibitem [{\citenamefont {Carrasco}\ and\ \citenamefont
  {Rodina}(2019)}]{Carrasco:2019qwr}%
  \BibitemOpen
  \bibfield  {author} {\bibinfo {author} {\bibfnamefont {John Joseph~M.}\
  \bibnamefont {Carrasco}}\ and\ \bibinfo {author} {\bibfnamefont {Laurentiu}\
  \bibnamefont {Rodina}},\ }\bibfield  {title} {\enquote {\bibinfo {title} {{UV
  considerations on scattering amplitudes in a web of theories}},}\ }\href@noop
  {} {\  (\bibinfo {year} {2019})},\ \Eprint {http://arxiv.org/abs/1908.08033}
  {arXiv:1908.08033 [hep-th]} \BibitemShut {NoStop}%
\bibitem [{\citenamefont {Carrasco}\ \emph {et~al.}()\citenamefont {Carrasco},
  \citenamefont {Rodina},\ and\ \citenamefont {Zekioglu}}]{hdHigherMult}%
  \BibitemOpen
  \bibfield  {author} {\bibinfo {author} {\bibfnamefont {J.~J.~M.}\
  \bibnamefont {Carrasco}}, \bibinfo {author} {\bibfnamefont {Laurentiu}\
  \bibnamefont {Rodina}}, \ and\ \bibinfo {author} {\bibfnamefont {Suna}\
  \bibnamefont {Zekioglu}},\ }\href@noop {} {}\bibinfo {note} {{\em In
  preperation.}}\BibitemShut {Stop}%
\bibitem [{\citenamefont {Bargheer}\ \emph {et~al.}(2012)\citenamefont
  {Bargheer}, \citenamefont {He},\ and\ \citenamefont
  {McLoughlin}}]{Bargheer2012gv}%
  \BibitemOpen
  \bibfield  {author} {\bibinfo {author} {\bibfnamefont {Till}\ \bibnamefont
  {Bargheer}}, \bibinfo {author} {\bibfnamefont {Song}\ \bibnamefont {He}}, \
  and\ \bibinfo {author} {\bibfnamefont {Tristan}\ \bibnamefont {McLoughlin}},\
  }\bibfield  {title} {\enquote {\bibinfo {title} {{New relations for
  three-dimensional supersymmetric scattering amplitudes}},}\ }\href {\doibase
  10.1103/PhysRevLett.108.231601} {\bibfield  {journal} {\bibinfo  {journal}
  {Phys. Rev. Lett.}\ }\textbf {\bibinfo {volume} {108}},\ \bibinfo {pages}
  {231601} (\bibinfo {year} {2012})},\ \Eprint {http://arxiv.org/abs/1203.0562}
  {arXiv:1203.0562 [hep-th]} \BibitemShut {NoStop}%
\bibitem [{\citenamefont {Johansson}\ and\ \citenamefont
  {Ochirov}(2015)}]{Johansson2014zca}%
  \BibitemOpen
  \bibfield  {author} {\bibinfo {author} {\bibfnamefont {Henrik}\ \bibnamefont
  {Johansson}}\ and\ \bibinfo {author} {\bibfnamefont {Alexander}\ \bibnamefont
  {Ochirov}},\ }\bibfield  {title} {\enquote {\bibinfo {title} {{Pure gravities
  via color-kinematics duality for fundamental matter}},}\ }\href {\doibase
  10.1007/JHEP11(2015)046} {\bibfield  {journal} {\bibinfo  {journal} {JHEP}\
  }\textbf {\bibinfo {volume} {11}},\ \bibinfo {pages} {046} (\bibinfo {year}
  {2015})},\ \Eprint {http://arxiv.org/abs/1407.4772} {arXiv:1407.4772
  [hep-th]} \BibitemShut {NoStop}%
\bibitem [{\citenamefont {Johansson}\ and\ \citenamefont
  {Ochirov}(2016)}]{Johansson:2015oia}%
  \BibitemOpen
  \bibfield  {author} {\bibinfo {author} {\bibfnamefont {Henrik}\ \bibnamefont
  {Johansson}}\ and\ \bibinfo {author} {\bibfnamefont {Alexander}\ \bibnamefont
  {Ochirov}},\ }\bibfield  {title} {\enquote {\bibinfo {title}
  {{Color-kinematics duality for QCD amplitudes}},}\ }\href {\doibase
  10.1007/JHEP01(2016)170} {\bibfield  {journal} {\bibinfo  {journal} {JHEP}\
  }\textbf {\bibinfo {volume} {01}},\ \bibinfo {pages} {170} (\bibinfo {year}
  {2016})},\ \Eprint {http://arxiv.org/abs/1507.00332} {arXiv:1507.00332
  [hep-ph]} \BibitemShut {NoStop}%
\bibitem [{\citenamefont {Johansson}\ and\ \citenamefont
  {Ochirov}(2019)}]{Johansson:2019dnu}%
  \BibitemOpen
  \bibfield  {author} {\bibinfo {author} {\bibfnamefont {Henrik}\ \bibnamefont
  {Johansson}}\ and\ \bibinfo {author} {\bibfnamefont {Alexander}\ \bibnamefont
  {Ochirov}},\ }\bibfield  {title} {\enquote {\bibinfo {title} {{Double copy
  for massive quantum particles with spin}},}\ }\href@noop {} {\  (\bibinfo
  {year} {2019})},\ \Eprint {http://arxiv.org/abs/1906.12292} {arXiv:1906.12292
  [hep-th]} \BibitemShut {NoStop}%
\bibitem [{\citenamefont {Chiodaroli}\ \emph {et~al.}(2017)\citenamefont
  {Chiodaroli}, \citenamefont {G\"unaydin}, \citenamefont {Johansson},\ and\
  \citenamefont {Roiban}}]{Chiodaroli2017ngp}%
  \BibitemOpen
  \bibfield  {author} {\bibinfo {author} {\bibfnamefont {Marco}\ \bibnamefont
  {Chiodaroli}}, \bibinfo {author} {\bibfnamefont {Murat}\ \bibnamefont
  {G\"unaydin}}, \bibinfo {author} {\bibfnamefont {Henrik}\ \bibnamefont
  {Johansson}}, \ and\ \bibinfo {author} {\bibfnamefont {Radu}\ \bibnamefont
  {Roiban}},\ }\bibfield  {title} {\enquote {\bibinfo {title} {{Explicit
  formulae for Yang-Mills-Einstein amplitudes from the double copy}},}\ }\href
  {\doibase 10.1007/JHEP07(2017)002} {\bibfield  {journal} {\bibinfo  {journal}
  {JHEP}\ }\textbf {\bibinfo {volume} {07}},\ \bibinfo {pages} {002} (\bibinfo
  {year} {2017})},\ \Eprint {http://arxiv.org/abs/1703.00421} {arXiv:1703.00421
  [hep-th]} \BibitemShut {NoStop}%
\bibitem [{\citenamefont {Du}\ \emph {et~al.}(2017)\citenamefont {Du},
  \citenamefont {Feng},\ and\ \citenamefont {Teng}}]{Du2017gnh}%
  \BibitemOpen
  \bibfield  {author} {\bibinfo {author} {\bibfnamefont {Yi-Jian}\ \bibnamefont
  {Du}}, \bibinfo {author} {\bibfnamefont {Bo}~\bibnamefont {Feng}}, \ and\
  \bibinfo {author} {\bibfnamefont {Fei}\ \bibnamefont {Teng}},\ }\bibfield
  {title} {\enquote {\bibinfo {title} {{Expansion of all multitrace tree level
  EYM amplitudes}},}\ }\href {\doibase 10.1007/JHEP12(2017)038} {\bibfield
  {journal} {\bibinfo  {journal} {JHEP}\ }\textbf {\bibinfo {volume} {12}},\
  \bibinfo {pages} {038} (\bibinfo {year} {2017})},\ \Eprint
  {http://arxiv.org/abs/1708.04514} {arXiv:1708.04514 [hep-th]} \BibitemShut
  {NoStop}%
\bibitem [{\citenamefont {Elvang}\ \emph {et~al.}(2019)\citenamefont {Elvang},
  \citenamefont {Hadjiantonis}, \citenamefont {Jones},\ and\ \citenamefont
  {Paranjape}}]{Elvang:2018dco}%
  \BibitemOpen
  \bibfield  {author} {\bibinfo {author} {\bibfnamefont {Henriette}\
  \bibnamefont {Elvang}}, \bibinfo {author} {\bibfnamefont {Marios}\
  \bibnamefont {Hadjiantonis}}, \bibinfo {author} {\bibfnamefont {Callum
  R.~T.}\ \bibnamefont {Jones}}, \ and\ \bibinfo {author} {\bibfnamefont
  {Shruti}\ \bibnamefont {Paranjape}},\ }\bibfield  {title} {\enquote {\bibinfo
  {title} {{Soft bootstrap and supersymmetry}},}\ }\href {\doibase
  10.1007/JHEP01(2019)195} {\bibfield  {journal} {\bibinfo  {journal} {JHEP}\
  }\textbf {\bibinfo {volume} {01}},\ \bibinfo {pages} {195} (\bibinfo {year}
  {2019})},\ \Eprint {http://arxiv.org/abs/1806.06079} {arXiv:1806.06079
  [hep-th]} \BibitemShut {NoStop}%
\bibitem [{\citenamefont {Low}\ and\ \citenamefont {Yin}(2019)}]{Low:2019ynd}%
  \BibitemOpen
  \bibfield  {author} {\bibinfo {author} {\bibfnamefont {Ian}\ \bibnamefont
  {Low}}\ and\ \bibinfo {author} {\bibfnamefont {Zhewei}\ \bibnamefont {Yin}},\
  }\bibfield  {title} {\enquote {\bibinfo {title} {{Soft Bootstrap and
  Effective Field Theories}},}\ }\href@noop {} {\  (\bibinfo {year} {2019})},\
  \Eprint {http://arxiv.org/abs/1904.12859} {arXiv:1904.12859 [hep-th]}
  \BibitemShut {NoStop}%
\bibitem [{\citenamefont {Carrillo-Gonzalez}\ \emph {et~al.}(2019)\citenamefont
  {Carrillo-Gonzalez}, \citenamefont {Penco},\ and\ \citenamefont
  {Trodden}}]{Carrillo-Gonzalez:2019aao}%
  \BibitemOpen
  \bibfield  {author} {\bibinfo {author} {\bibfnamefont {Mariana}\ \bibnamefont
  {Carrillo-Gonzalez}}, \bibinfo {author} {\bibfnamefont {Riccardo}\
  \bibnamefont {Penco}}, \ and\ \bibinfo {author} {\bibfnamefont {Mark}\
  \bibnamefont {Trodden}},\ }\bibfield  {title} {\enquote {\bibinfo {title}
  {{Shift symmetries, soft limits, and the double copy beyond leading
  order}},}\ }\href@noop {} {\  (\bibinfo {year} {2019})},\ \Eprint
  {http://arxiv.org/abs/1908.07531} {arXiv:1908.07531 [hep-th]} \BibitemShut
  {NoStop}%
\bibitem [{\citenamefont {Bern}\ \emph
  {et~al.}(2019{\natexlab{c}})\citenamefont {Bern}, \citenamefont
  {Parra-Martinez},\ and\ \citenamefont {Sawyer}}]{Bern:2019wie}%
  \BibitemOpen
  \bibfield  {author} {\bibinfo {author} {\bibfnamefont {Zvi}\ \bibnamefont
  {Bern}}, \bibinfo {author} {\bibfnamefont {Julio}\ \bibnamefont
  {Parra-Martinez}}, \ and\ \bibinfo {author} {\bibfnamefont {Eric}\
  \bibnamefont {Sawyer}},\ }\bibfield  {title} {\enquote {\bibinfo {title}
  {{Non-renormalization and operator mixing via on-shell methods}},}\
  }\href@noop {} {\  (\bibinfo {year} {2019}{\natexlab{c}})},\ \Eprint
  {http://arxiv.org/abs/1910.05831} {arXiv:1910.05831 [hep-ph]} \BibitemShut
  {NoStop}%
\end{thebibliography}%

\end{document}